\input{psfig}

\documentstyle[iopconf1]{article}
\begin{document}
\title{ Non-universal Soft SUSY Breaking Effects 
on Dark Matter and on Physics at Colliders}

\author{Pran Nath\dag\ and R. Arnowitt\ddag}

\affil{\dag\ Department of Physics, Northeastern University, 
	Boston, MA, 02115, USA}

\affil{\ddag\ Center for Theoretical Physics, Department of Physics,
Texas A\&M University, College Station, TX 77843-4242, USA}

\beginabstract
Implications of the effects of non-universalities of the soft SUSY
breaking parameters on dark matter and on physics at colliders is 
discussed.  
\endabstract

\section{Introduction}
In this paper we review the effects of non-universalities of soft 
SUSY breaking parameters on low energy physics.  Most of the existing
analyses of SUSY phenomenology has been within the framework of 
minimal SUGRA unification\cite{cham1,applied} which is  parametrized in 
terms of four
parameters under the constraint of radiative breaking of the
electro-weak symmetry. These can be taken to be the universal
scalar mass $m_0$, the gluino mass $m_{\tilde g}$, the trilinear coupling
$A_t$, and tan$\beta=<v_2>/<v_1>$, where $v_2$ gives mass to the up quark and
$<v_1>$ gives mass to the down quark. 
 However, the framework of supergravity
and string theory allows for non-universalities of the soft
SUSY breaking parameters to appear\cite{soni,planck}.
 There are two main sources  which can give rise to 
non-universalities of  this type  in supergravity 
unified models. One  of these is the gauge kinetic energy function
$f_{\alpha\beta}$ while the second is a non-flat generation dependent
Kahler potential.  A general gauge kinetic
function introduces  non-universalities in the gaugino masses 
 while a non-flat Kahler potential 
leads to non-universal scalar masses 
at the scale where SUSY breaks. One of  the purposes of the
analysis discussed in this paper is  to identify signatures of such
non-universalities in low energy physics, such as in dark matter and
in physics at colliders.

		The outline of the paper is as follows. In Sec.2
		we give a general discussion of non-universalities.
		In Sec.3 we discuss the implications of non-universalities
		on dark matter. In Sec.3 we discuss how non-universality
		effects can be discerned from precision analyses of the 
		sfermion mass spectrum.

\section{Non-universalities  in Supergravity Unification}
We discuss now the two main sources of non-universalities in 
supergravity unified models. One of the sources is the 
gauge kinetic energy function 
 $f_{\alpha\beta}$
which in general can possess Planck scale corrections so that\cite{hill,das}
 \begin{equation}
  f_{\alpha\beta}= \delta_{\alpha \beta}+ 
 \frac{c}{2M_{Planck}} f_{\alpha\beta \gamma} \Sigma^{\gamma}
 \end{equation}
 where  $\Sigma$ is the adjoint Higgs. The gauge kinetic energy function
 also  enters in the gaugino masses and one gets 
 
  \begin{equation}
(m_{\frac{1}{2}})_{\alpha\beta}= \kappa^{-1}\langle
G^a (K^{-1})^a_b~Ref_{\alpha\beta, b}^{\dag}\rangle m_{3/2}
\end{equation}	
\noindent
where $\kappa=1/M_{Planck}$,
 $G=\kappa^2 K+\ell n [\kappa^6\mid W \mid^2]$, K is the Kahler 
potential, W is the superpotential, 
$G^a\equiv\partial G/\partial Q_a$ and($K^{-1})^a_b$ is the matrix
inverse of the Kahler metric $K_b^a$. After spontaneous breaking of 
the GUT symmetry one generates non-universalities of the gaugino masses
 of size O($M/M_{Pl}$) so that\cite{das}

 \begin{equation}
 M_i=\frac{\alpha_i(Q)}{\alpha_G}(1+c'\frac{M}{M_{pl}}n_i) m_{\frac{1}{2}}
 \end{equation}
 where $\alpha_G$ is the GUT coupling constant, M is the GUT scale, 
$\alpha_i$ are the subgroup gauge coupling constants, and 
$n_i=(2,-3,-1)$ for the subgroups ($SU(3)$, $SU(2)$, $U(1)$), and 
$c'$ while proportional to c is an independent parameter. 

	There may already be experimental evidence for the presence of
	the Planck scale correction of the type discussed above. Thus
	the renormalization group analyses of the gauge coupling constants 
	within the minimal SU(5) SUSY/SUGRA unification show that $\alpha_s$
	is about $2\sigma$ higher than the current world average. This
	situation can be corrected by an inclusion of the Planck scale 
	correction with c  
	$\sim 1$. Of course a non-vanishing c term also generates a 
	correction to the gaugino masses as seen above and such corrections
	affect scaling relations. Thus in the absence of the $c'$ term one
	finds that over most of the parameter space one has the scale
	relation\cite{scaling} 

\begin{eqnarray}
2m_{\chi_{1}}^0&\cong& m_{\chi_{1}}^{\pm}\cong m_{\chi_{2}}^0\simeq {1\over 3}
m_{\tilde g}\nonumber\\
m_{\chi_{3}}^0 &\cong& m_{\chi_{4}}^0 \cong m_{\chi_{2}}^{\pm}\simeq \mu >>
m_{\chi_{1}}^0\nonumber\\
m_{A}&\cong& m_{H^0} \cong m_H^{\pm} 
\end{eqnarray}
arising because  of the fact that over most of the parameter space of the
SUGRA unfiied models one has $\mu^2/M_Z^2>> 1$. Some of these scaling
laws break down when non-universalities are included;  e.g., one
gets\cite{das} 
$2m_{\chi_{1}}^0$ $\neq m_{\chi_{1}}^{\pm}$ $\cong $ 
$m_{\chi_{2}}^0$ $\neq {1\over 3} m_{\tilde g}$.
From Eq.(3) we see that for $M/M_{Pl}\approx 1/50$, $c'\sim 3$, one can get
corrections to the gaugino masses of O(20\%) generating a splitting  
of O(30\%) in the ratio of the gaugino masses from their $c'=0$ values because
$n_i$ do not have fixed sign for all i.
Such effects could be seen in accurate measurements of  the mass
spectra of charginos, neutralinos, and gluinos (see Sec.4 for 
discussion of the accuracy with which mass spectra can be measured in
the future). Further, in the deep scaling region, i.e., if $(|\mu/M_Z|>5)$, 
 scaling is expected
to hold to a few \% accuracy  when $c'=0$. Thus in this region 
significant contributions from $c'$, i.e., if $c'\sim 2-3$ could be visible
as deviations from scaling.
 Next we discuss non-universalities in the scalar soft SUSY breaking sector.
 As pointed out earlier 
 these arise due to the presence of  a general Kahler potential
 which can be expanded in terms of the  visible sector fields ($Q_a, Q^a$)
 as follows\cite{soni,planck}
	\begin{equation}
	K=\kappa^2K_0+K^a_b Q_aQ^b+(K^{ab}Q_a Q_b+ h.c.)+..
	\end{equation}
	where $K_0, K_b^a, ..$ etc are in general functions of the fields
	in the hidden sector. For the minimal suprgravity unification case
	the assumption  $K^a_b=K(h,h^{\dagger})\delta^a_b$,
	 where h are the fields in the hidden sector, 
	which leads to universality when SUSY breaks. 
	Non-universalities appear when one gives up this assumption.
	However, one cannot allow an arbitrary set of non-universalities
	in the soft SUSY breaking sector because of the stringent
	experimental  constraints on flavor changing neutral currents 
	(FCNC). One sector where the FCNC constraints are not so 
	stringent is the Higgs sector. Thus in this sector one may 
	phenomenogically parametrize the non-universalities in the
	following way\cite{matallio,olech,polon,berez,nonuni} 
	
 \begin{equation}
m_{H_1}^2 =  m_0^2 (1 + \delta_1),  ~~m_{H_2}^2 = m_0^2 (1 + \delta_2)
\end{equation}
where one limits the $\delta_i$ so that $|\delta_i|\leq 1$ (i=1,2). However,
it was pointed out in ref.\cite{nonuni} that the non-universalities in
the Higgs sector and in the third generation sector are strongly coupled.
Thus one should also include non-universalities in the third generation,
i.e.,   
\begin{equation}
  m_{\tilde Q_L}^2=m_0^2(1+\delta_3), ~~m_{\tilde U_R}^2=m_0^2(1+\delta_4)
\end{equation}  
where as before one limits $|\delta_i|\leq 1$(i=3,4). 

	One of the ways non-universalities can affect low energy physics
	is via their effects on $\mu^2$. One may in general write
$\mu^2=  \mu_0^2+  \Delta\mu^2$,
where $\mu_0^2$ is the part for universal soft SUSY
breaking and $\Delta\mu^2$ is the correction that arises due to 
non-universalities. For tan$\beta$ small enough one can neglect the
b quark coupling and obtain the following analytic expression for 
$\Delta\mu^2$\cite{nonuni} 
\begin{equation}
\Delta\mu^2=m_0^2
\frac{1}{t^2-1}(\delta_1-\delta_2t^2-\frac{ D_0-1}{2}(\delta_2 +
\delta_3+\delta_4)t^2)+ \frac{3}{5}\frac{t^2+1}{t^2-1}S_0p
\end{equation}
where  t$\equiv tan\beta$, 
$D_0=1-(\frac{m_t}{m_f})^2), ~~m_f\simeq 200 sin\beta ~GeV$, 
$S_0=Tr(Ym^2)$, and p=0.0446.
The $S_0$ term is the anomaly term which vanishes for the universal  
case because of anomaly cancellation, i.e., Tr($Ym^2$)=0. 

Part of the mass spectrum which is affected sensitively by 
non-universalities is the spectrum of  the third generation masses,
 for example, the stop masses which are governed by the matrix 
\begin{equation}
\left(
{{ {m_{\tilde t_L}^2}\atop{-m_{t} (A_t + \mu ctn \beta)}}
{{-m_t (A_t +\mu ctn \beta)} \atop {m_{\tilde t_R}^2}   }}
\right)
\end{equation}
Non-universalities enter via corrections to  $\mu$ in the off-diagonal 
elements and corrections to the diagonal elements so that 
$m^2_{\tilde t_{L}}$ = $m^2_{\tilde t_{L}}(0)$ + $\Delta m^2_{\tilde t_{L}}$,
where  $m^2_{\tilde t_{L}}(0)$ is the part for universal 
 soft SUSY breaking and $\Delta m^2_{\tilde t_{L}}$ is the 
non-universality correction\cite{nonuni}  
\begin{equation}
\Delta m^2_{\tilde t_{L}} = m_0^2(\frac{(D_0-1)}{6}(\delta_2+\delta_3+\delta_4)
+\delta_3) 
\end{equation}
Similarly 
$m^2_{\tilde t_{R}}$ = $m^2_{\tilde t_{R}}(0)$ + $\Delta m^2_{\tilde t_{R}}$,
where $m^2_{\tilde t_{R}}(0)$ is the universal part and 
$\Delta m^2_{\tilde t_{R}}$ is given by\cite{nonuni}

\begin{equation}
\Delta m^2_{\tilde t_{R}} = m_0^2(\frac{(D_0-1)}{3}(\delta_2+\delta_3+\delta_4)
+\delta_4)
\end{equation}
The above analysis shows that  $\delta_3$ and $\delta_4$ enter on an equal 
footing with $\delta_1$ and $\delta_2$ so the non-universalities in the
Higgs sector and in the third generation are strongly coupled as
stated earlier.
 
\section{Effects of Non-universal Soft SUSY Breaking on Dark Matter}
In this section we discuss the effects of non-universalities on dark
matter. We review first the basic elements of the analysis. 
The analysis assumes R parity invariance, and it can be shown that 
over most of the parameter space of the model the lightest neutralino is 
 also the lowest supersymmetric particle (LSP)\cite{scaling} 
 and hence a candidate for cold dark matter (CDM). 
 For the purpose of the analysis here we
 shall impose the dark matter constraint  

\begin{equation}
0.1\leq \Omega_{\tilde\chi_1^0}~h^2\leq 0.4 
\end{equation}
where $\Omega_{\tilde\chi_1^0}$ in the ratio of the mass density of the
LSP relic to  the critical mass density needed to close the universe. The
quantity that can be computed  theoretically is 
$\Omega_{\tilde\chi_1^0}h^2$, where 
 h is the Hubble parameter in units of 100 km/sMpc,
 and is given by\cite{jungman} 

\begin{equation}
\Omega_{\tilde\chi_1^0} h^2\cong 2.48\times 10^{-11}{\biggl (
{{T_{\tilde\chi_1^0}}\over {T_{\gamma}}}\biggr )^3} {\biggl ( {T_{\gamma}\over
2.73} \biggr)^3} {N_f^{1/2}\over J ( x_f )}
\end{equation}
~\\
 Here $x_f= kT_f/m_{\tilde{\chi}_{1}}$, where $T_f$ is the freezeout
 temperature, $N_f$ is the number of degrees of freedom at 
 freezeout,  $(T_{\tilde\chi_1^0}/T_{\gamma})^3$ is  the reheating
 factor, $ T_{\gamma}$ is the current micro-wave background 
 temperature and  $J~ (x_f)$ is given by
~\\
\begin{equation}
J~ (x_f) = \int^{x_f}_0 dx ~ \langle~ \sigma \upsilon~ \rangle ~ (x) GeV^{-2}
\end{equation}
~\\
\noindent
where $\sigma$ is the annihilation cross-section for the neutralinos, $v$ is 
their relative velocity and  $<\sigma v>$ is the thermal average. In
the computation of the thermal average we have used the accurate 
method\cite{greist,accurate}. 
         
         There are many techniques discussed in the literature for the
         detection of dark matter. One  interesting possibility
         is the direct detection via scattering of neutralinos off
         nuclei. This process in governed by the basic 
         interaction\cite{goodman,bottino,bednyakov,an2,na2}  
\begin{equation}
L_{eff}= ({\bar{\tilde\chi}_1}{\gamma^{\mu}}\gamma_5\tilde\chi_1)[\bar{q}\gamma_{\mu}
(A_LP_L+A_RP_R)q]+(\bar{\tilde\chi}_1 \tilde\chi_1)(\bar{q}C m_qq) 
\end{equation} 
which consists of a spin dependent interaction governed by $A_L$
and $A_R$ terms and a scalar interaction governed by the  C term. 
The event rates are given by\cite{goodman}  

\begin{equation}
R=\left[ R_{SI}+R_{SD}\right ] \left [{\rho_{\tilde{\chi}_{1}}\over 0.3GeV
cm^{-3}}\right ] \left [{v_{\tilde{\chi}_{1}}\over 320 km/s}\right ]{events\over
kg~ da}  
\end{equation}
\noindent
where $ R_{SD}$ is the spin dependent part, $R_{SD}$ is the spin 
independent part, $v_{\tilde{\chi}_{1}}$ is the velocity of relic neutralinos
in our galaxy impinging on the target,  and  $\rho_{\tilde\chi_1}$ 
is the local density of the relic neutralinos. $R_{SD}$ is given by
     
\begin{equation}
R_{SD} = {16 m_{\tilde{\chi}_{1}}M_N\over \left [M_N+m_{\tilde{\chi}_{1}}\right ]^2}
\lambda^2J(J+1)\mid A_{SD}\mid^2
\end{equation}
\noindent
where  M$_N$ is the mass and J is the spin 
of the target nucleus, $\lambda$ is defined
so that 
$<N\mid\sum
{\stackrel{\rightarrow}{S}_i}\mid N>$=$\lambda<N\mid{\stackrel{\rightarrow}
{J}}\mid N>$, and $A_{SD}$ is the spin dependent amplitude.
Similarly  $ R_{SI}$  is given by  

\begin{equation}
R_{SI}={16m_{\tilde{\chi}_{1}}M_N^3M_Z^4\over\left
[{M_N+m_{\tilde{\chi}_{1}}}\right ]^2}{\mid A_{SI}\mid^2}  
\end{equation}
where  $A_{SI}$ is the spin independent amplitude.  For heavy targets
one has  $R_{SD}\sim 1/{M_N}$ and  
 R$_{SI}\sim M_N$. Thus for heavy targets one expects that the scalar 
 interaction will eventually dominate over the spin dependent interaction.

		We discuss now the effect of the  Higgs sector 
		 non-universalities when $\delta_3=0=\delta_4$. We consider
		 three cases for comparison
(i)$\delta_1$ =0=$\delta_2$,~~
(ii)$\delta_1$ =-1=-$\delta_2$,~~
(iii)$\delta_1$=1=-$\delta_2$. 
The result of the analysis is  exhibited in Fig.1. 
Here we find that for the $\delta_1=-1=-\delta_2$ case $\Delta \mu^2$
receives a negative contribution which tends to
reduce $\mu^2$ which raises event rates. It also 
 drives $\mu^2$ towards
the tachyonic limit eliminating part of the parameter space. Typically
the part of the parameter that gets eleiminated for 
$m_{\chi_1}<65$ GeV is the small $tan\beta$ region. Elimination of small
tan$\beta$ tends to drive the minimum of the event rate higher
which is what is seen in Fig.1. For the region $m_{\chi_1}>65$ GeV
Landau pole and $m_{1/2}^2$ term  
effects are the dominant terms so the effects of 
non-universalities here are somewhat suppressed. 
 For the case $\delta_1=1=-\delta_2$  
the effect is opposite to that for  the previous case. A similar analysis
holds for the effects of non-universalities in the third generation
sector with $\delta_3$ and $\delta_4$ acting opposite to $\delta_2$.
 From Fig.1 it is seen that the event rates lie 
in a wide range O(1-$10^{-5}$)
event/kgd. The sensitivity of  current detectors is rather 
limited\cite{bernabei,bottino2}, and one needs more sensitive 
detectors\cite{cline} to probe a majority of the parameter space of 
 supergravity models.

\begin{figure*}
\begin{center}
\hspace*{0.3in}
\psfig{figure=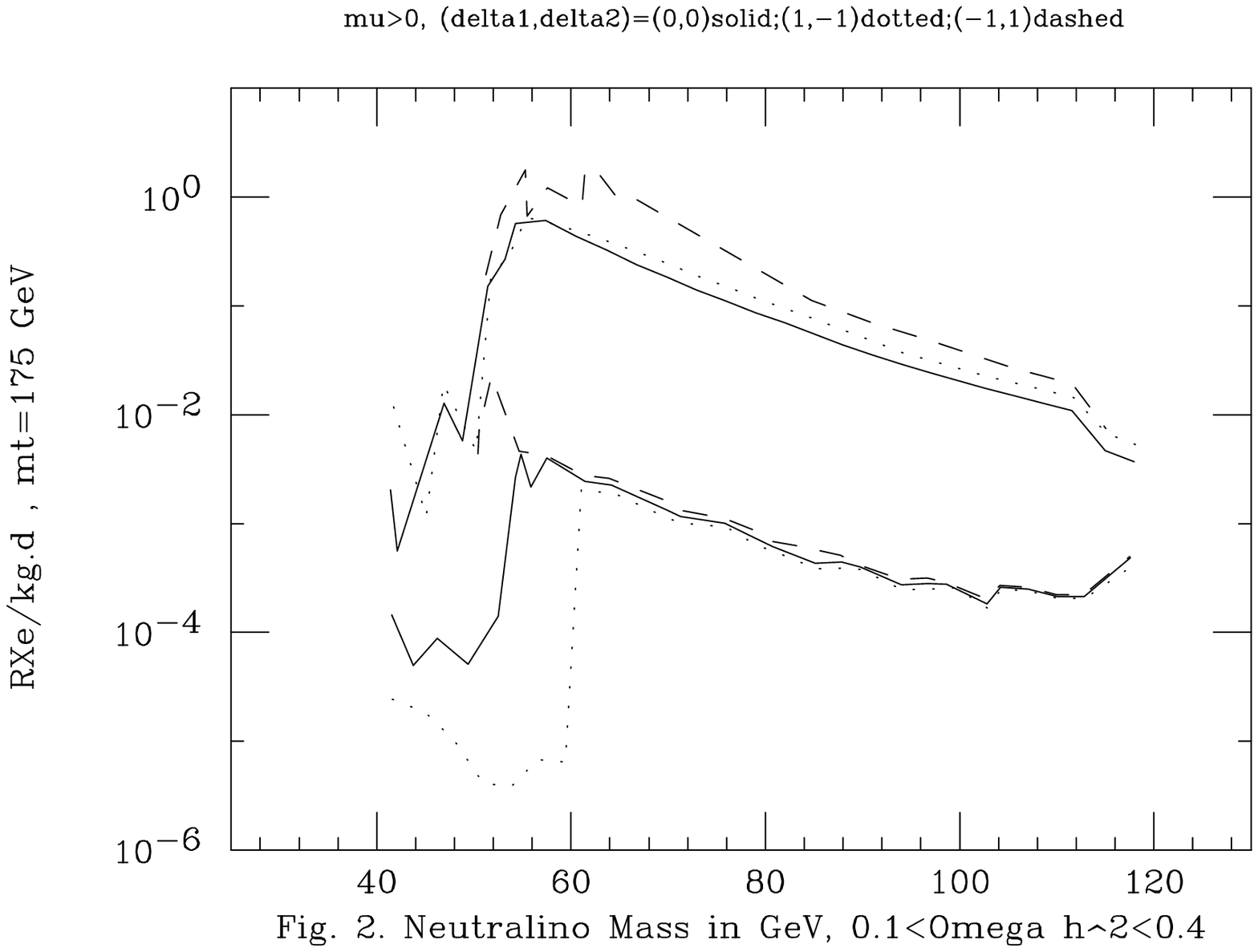,height=1.5in}
\end{center}
\caption[]{ 
Maximum and minimum of event rates/kg d for xenon  for $\mu>0$
for the case
 when $\delta_3=0=\delta_4$ and 
 (a)$\delta_1=0=\delta_2$(solid), (b)$\delta_1=1=-\delta_2$(dotted), 
 and (c)$\delta_1=-1=-\delta_2$ (dashed)
  when $0.1<\Omega_{\tilde\chi_1} h^2<0.4$, and $m_t$=175 GeV.
  (From Ref.\cite{nonuni}).}
\label{fig:1}
\end{figure*}
Finally,  we consider the  effects 
more accurate determinations of the cosmological parameters
by future satellite experiments, such as MAP and PLANCK,
 will have on dark matter analyses. It is expected that these experiments
 will determine the cosmological parameters to within (1-10$\%$) 
 accuracy\cite{kosowsky,dodelson}.
 There are a variety of models which fit the current cosmological data,
 such as $\Lambda CDM, \nu CDM$, etc. For illustration we consider 
 $\Lambda CDM$ with the parameters $\Omega_{CDM}=0.4$,   $\Omega_{B}=0.05$,  
  $\Omega_{\Lambda}=0.55$, and h=0.62 which give a reasonable fit to the
  current  astro-physical data. Assuming that  PLANCK reaches its expected 
   accuracy\cite{kosowsky} one  would find the constraint
   \begin{equation}
   \Omega_{CDM} h^2=0.154\pm 0.017
   \end{equation}
   The implications of this constraint on event rate analyses is 
   exhibited in Fig.2. One finds two generic features valid for  both the
   universal as well as for the non-universal case. The first is that
   the  maximum and the minimum corridors appear to shrink giving a narrower
   range in which the event rates can lie. The second more potent result
   is that the  new $\Omega h^2$ constraint implies that the gluino mass
   should lie below 520 GeV. An analysis anologous to Fig.2 using a  $2\sigma$ 
   error corridor increases the upper limit on the gluino mass to 560 GeV.
    Much of this gluino mass domain  can be probed with an upgraded Tevatron 
    which can probe a 
   significant part of the above parameter space in the gluino 
   mass, i.e., up to about 450 GeV, with an integrated luminosity of 
   about $25fb^{-1}$\cite{kamon,tev2000}.

\begin{figure*}
\begin{center}
\hspace*{0.3in}
\psfig{figure=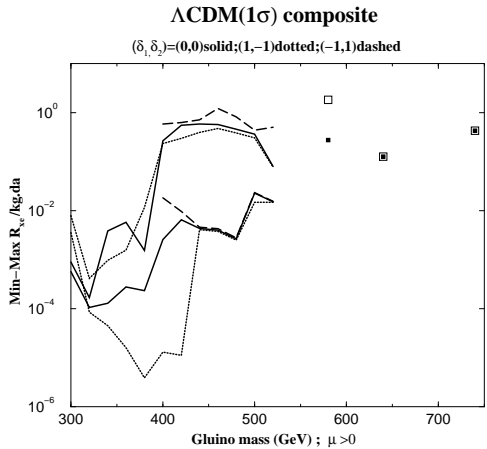,height=2.4in}
\end{center}
\caption[]{Same as Fig.1 except that the  constraint 
$\Omega h^2=0.154\pm 0.017$ is imposed. The isolated points are for the
$\delta_1=-1$, $\delta_2=1$ case.}
\label{fig:2}
\end{figure*}

\section{Sfermion Mass Spectrum as a Probe of  Non-universalities}
The use of sfermion spectrum as a probe of different patterns of GUT symmetry 
 breaking has already been discussed in the literature\cite{kawamura}. 
We discuss here the interesting possibility that the non-universalities
of the soft SUSY breaking can affect the sfermion masses and thus a precision
measurement of such masses would act as a probe for the existence of 
non-universalities\cite{planck}.
Thus assume, for example, the existence of non-universalities at 
the GUT scale in an SU(5) invariant theory. Then in this case one would
find that the mass differences 
	$m_{\tilde u_L}^2$ - $m_{\tilde u_R}^2$,
	$m_{\tilde u_L}^2$ - $m_{\tilde u_R}^2$,
	$m_{\tilde u_L}^2$ - $m_{\tilde u_R}^2$,
at the electro-weak scale would still be independent of the non-universalities. Further, by evolving
the sfermion masses beyond the GUT scale one can determine if the different
scalar masses unify at a common scale. If they do then the unification 
would attest to the existence of a universal soft SUSY breaking at the
SUSY breaking scale which could be the string scale and at the same 
time allow us to experimentally deduce the value of the string scale.
	
		These ideas can be easily extended to other choices of
		the GUT group, such as SO(10), SU(3)$^3$, $G_{SM}$, etc. 
		We consider the SO(10) case and  assume that
		SO(10) breaks in one step at the scale $M_{GUT}$ down to the
		Standard Model gauge group $G_{SM}$. Here because
		of rank reduction one has  D term contributions 
		to the matching conditions at $M_{GUT}$\cite{drees,kawamura}. 
		Assuming that
		the matter spectrum falls in the 16-plets of SO(10) 
		which decomposes into 16=10+$\bar 5$+1 of SU(5), and the
		$5+\bar 5$ of Higgs lies in the 10 of SO(10) 
		one can define 
	\begin{equation}
	m_5^2=\tilde m_0^2(1+\delta_5),
	m_{H_1}^2=\tilde m_0^2(1+\delta_1),
	m_{H_2}^2=\tilde m_0^2(1+\delta_2),
	\end{equation}		
where we have  absorbed  $\delta_{10}$ into the definition of the mass of 
the 10 of SU(5). 
In this case the matching conditions imply 
$\delta_5$ = $\delta_1$ - $\delta_2$. This constraint can be tested if 
one has accurate measurements of sfermion masses. 
 One expects that measurements accurate to 
a few percent  may be possible in future 
colliders\cite{baer,hinch,tsukamoto,feng,snowmass}, which  			
  would  determine the non-universalities at the 
$\sim 10\%$ level allowing one to test physics in the post GUT
region up to the string scale\cite{planck}.

\section*{Acknowledgements}
 This work was  supported in part by NSF grants PHY-9602074 and
 PHY-9722090.

\end{document}